\documentclass[sigconf]{acmart}

\AtBeginDocument{%
  }

\copyrightyear{2023}
\acmYear{2023}
\setcopyright{acmcopyright}\acmConference[SAC '23]{The 38th ACM/SIGAPP Symposium on Applied Computing}{March 27-March 31, 2023}{Tallinn, Estonia}
\acmBooktitle{The 38th ACM/SIGAPP Symposium on Applied Computing (SAC '23), March 27-March 31, 2023, Tallinn, Estonia}
\acmPrice{15.00}
\acmDOI{10.1145/3555776.3577623}
\acmISBN{978-1-4503-9517-5/23/03}

\begin{document}

\title{\textit{EdGCon}: Auto-assigner of Iconicity Ratings Grounded by Lexical Properties to Aid in Generation of Technical Gestures}

\renewcommand{\shorttitle}{ASL Technical Iconicity Rating}

\author{Sameena Hossain, Payal Kamboj, Aranyak Maity, Tamiko Azuma, Ayan Banerjee, Sandeep K. S. Gupta}
\email{{shossai5, pkamboj, amaity1, Tamiko.Azuma, abanerj3, Sandeep.Gupta}@asu.edu}

\orcid{0000-0001-6288-0988}

\affiliation{%
	\institution{IMPACT Lab, (\texttt{impact.lab.asu.edu}), Arizona State University}
	\city{Tempe, Arizona}
	\country{USA}
}

\renewcommand{\shortauthors}{Hossain et al.}

\begin{abstract}
  Gestures that share similarities in their forms and are related in their meanings, should be easier for learners to recognize and incorporate into their existing lexicon. In that regard, to be more readily accepted as standard by the Deaf and Hard of Hearing community, technical gestures in American Sign Language (ASL) will optimally share similar in forms with their lexical neighbors. We utilize a lexical database of ASL, ASL-LEX, to identify lexical relations within a set of technical gestures. We use automated identification for 3 unique sub-lexical properties in ASL- \textit{location}, \textit{handshape} and \textit{movement}. $EdGCon$ assigned an iconicity rating based on the lexical property similarities of the new gesture with an existing set of technical gestures and the relatedness of the meaning of the new technical word to that of the existing set of technical words. We collected 30 ad hoc crowdsourced technical gestures from different internet websites and tested them against 31 gestures from the DeafTEC technical corpus. We found that $EdGCon$ was able to correctly auto-assign the iconicity ratings 80.76\% of the time.\footnote{This is a preprint version of conference paper accepted for publication in ACM SAC 2023.}
  
\end{abstract}




\maketitle

\section{Introduction}
With continued focus on accessibility in education in recent years, total Deaf and Hard of Hearing (DHH) student enrollment in STEM courses at 4-year undergraduate colleges ($\approx$ 17\%) has become nearly the same as hearing student enrollment ($\approx$ 18\%). However, only 0.19\% of DHH students pursue postgraduate education as opposed to nearly 15\% of hearing individuals~\cite{GD}. Only 20\% of deaf people between the ages of 18 to 44 attend post-secondary educational institutions each year~\cite{NCES}, and only a small subset will enroll in technical courses. This disparity results in DHH individuals having reduced access to higher-level skilled jobs in the technological fields that require postgraduate education and offer up to 31\% higher salaries~\cite{NITDRIT}.

One of the biggest hurdles in technical education for the DHH population is communicating specific technical terms through gestures. We define \textbf{technical words} as- a) words that are not standard in American Sign Language (ASL) but are commonly used in the technical field to identify various components or procedures, and b) words that are used in the technical field and are included within the current ASL lexicon, but the signed meaning denotes a concept that is different from the meaning used in the technical context. Frequently, these technical words must be fingerspelled, which is time consuming, especially for longer words. There have been several initiatives to generate a technical sign corpus for computer science (CS), including $ASLClear$~\cite{ASLClear} and $DeafTec$~\cite{DeafTec}. These efforts have led to the development of a repository of CS technical gestures, and are designed to facilitate technical and academic training of DHH students. Although such initiatives are significant steps towards a solution, there are several problems: a) the repository is a non-curated collection of gestures, enacted by several participants, thus the same technical word can have multiple different gesture representations, b) many technical words are still fingerspelled, and c) there is no effort to identify similarities in these technical gestures and to utilize these similarities to aid in generation of new technical gestures that can be easily recognized and adopted by learners. 

We hypothesize that for faster adoption and recognition by learners, any new gesture should be based on the $lexicon$ of a sign language and should convey the meaning of the word. In ASL-LEX (\cite{ASL-LEX}), form components of an ASL gesture are defined as the \textit{\textbf{Sub Lexical Properties}} and how much the gesture visually matches its meaning as the \textit{\textbf{Iconicity}}. Gestures that are similar in form are identified to be in \textit{\textbf{lexical neighborhoods}}. In recent years, different aspects of \textit{\textbf{Iconicity}} in gesture-speech recognition have been the topic of linguistics and educational research. Iconic gestures have been identified to be helpful in speech comprehension by multiple research studies surveyed in \cite{IconicitySpeechComprehension}. Iconic gestures also contribute to better understanding and increased word production in school aged children (\cite{IconicityWordProduction}). \textit{\textbf{Iconicity}}, hence, can also play a vital role in supplementing spoken vocabulary development with gestural communication in children with autism, Down Syndrome and Typical Development (TD)(\cite{GestureDownSyndrome}).

In this work, we the  \textit{\textbf{EdGCon}} tool that utilizes the concepts of \textit{iconicity} and \textit{sub lexical properties} \cite{ASL-LEX} to automatically assign \textbf{\textit{iconicity ratings}} to technical gestures. $Iconicity$, as utilized in gesture-speech research, is \textbf{subjective}. In ASL-LEX, ASL gestures are assigned $iconicity$ ratings on a scale of 1 to 7 (1 for ``least iconic'' and 7 for ``most iconic'') based on the observations of hearing individuals. This $subjectivity$ poses a challenge in automatic assignment of \textit{iconicity ratings} to newly generated technical gestures. To ensure objectivity of the \textit{iconicity ratings}, we follow a two step process which is the main contribution of this paper.

\subsection{Contributions \& Constraints}

We present a tool to automatically assign \textbf{\textit{Iconicity Ratings}} by identifying the lexical properties of a gesture and assessing how closely the gesture relates to the meaning of the word. For a \textbf{newly generated gesture} of a technical word (\textbf{\textit{step 1}}), we first identify the similarities in the form of the \textbf{new gesture} and the gestures in the existing technical corpus. Closest neighbor is identified based on the similarities in their form. Utilizing \textit{GloVe} \cite{GloVe}, we find the similarities between the technical word (corresponding to the new gesture) and the closest neighbor word (corresponding to the closest neighbor gesture). In \textbf{\textit{step 2}}, we compare the similarity with a predetermined threshold to assign iconicity rating to the newly generated gesture. If the similarity identified in step 1 is above the set threshold, it indicates that the technical word is similar in meaning to the closest neighbor word and that the corresponding gestures are also similar in their forms. The newly generated gesture can be considered as iconic as its closest neighbor, and we can assign the same iconicity rating to the newly generated gesture. 

We have designed the process abiding by three constraints to adhere to the ASL lexicon- a) We do not introduce new handshapes because we want new gestures to be consistent with the established ASL Lexicon, b) We want to preserve iconicity where possible (iconic signs are preferred), and c) Words with conceptual similarity are expected to have high neighborhood proximity with each other.

\section{Preliminaries}

Structure and Grammar of ASL have been studied since 1960 (\cite{LiddelASLGrammar,StokoeStructure,stokoe1970study,stokoe1976dictionary,stokoe2005sign}). Considering the ad hoc nature of how ASL gestures have been created traditionally, these studies are rich resources that can be utilized in further research into the relationship between the components of ASL gestures and their meaning. 


There have also been efforts to develop interconnected network of ASL gestures that are similar in form (\cite{ASLNet2019,ASL-LEX,HossainSemCog}) following the precedence of the network of interconnected English words based on their meanings and concepts \cite{WordNet, GloVe}. In this section, we discuss the tools and concepts used in this paper.

\subsection{ASL-LEX}
\label{ASLLEX}
ASL-LEX \cite{ASL-LEX} is a lexical database of American Sign Language (ASL) where ASL gestures are listed with their different lexical properties. This lexical information is collected based on the observations of hearing and DHH individuals, which are further discussed below.

\noindent\textit{Sign Identification}- English word with canonical meaning of the sign.

\noindent\textit{Frequency}- How often they felt the sign appears in everyday conversation.

\noindent\textit{Iconicity}- If the visual properties of sign are related to the meaning of the word, i.e. how much the sign looks like what it means. Hearing individuals use a 1 to 7 scale of rating to rate how much the gesture looks like its meaning- 7 indicating the gesture is most iconic (looks exactly like what it means) and 1 indicating the gesture is least iconic (does not look like what it means).

\noindent\textit{Lexical Information}- Nouns, Verbs, Adjectives, Adverbs \& closed -class items (conjunction, preposition, pronoun, interjection). Lexical class of a sign depends on the context it is used in.

\noindent\textit{Sign Length \& Clip Length}- \textbf{Sign onset} is the first video frame in which fully formed handshape contacted the body and \textbf{offset} is the last video frame in which hand contacted the body.

\noindent\textit{Phonological Coding or Sub-lexical Properties}- Sign Type (One handed, two handed, symmetrical or alternating), Location (Major Location \& Minor Location), Selected Fingers (Group of fingers that move), Flexicon (1 of nine degrees of flexicon). \textbf{Handshapes} are defined as unique combinations of \textit{Selected fingers} \& \textit{flexicon}.

\noindent\textit{Movement}- Path movement of the dominant hand through x y z space.

\noindent\textit{Neighborhood Density}- Maximal Neighborbhood Density (Signs that share any 4 of the 5 sub-lexical properties), Minimal Neighborhood Density (Signs that overlap with at least one feature of any kind with the target) and Parameter-based Neighborhood Density (Major Location, Movement, Selected Fingers and Flexicon).

\noindent\textit{Sub-lexical Frequency}- Count of the number of signs that are specified for that phonological property.

Our proposed tool \textbf{\textit{EdGCon}}, utilizes the \textbf{\textit{Sub-lexical Properties}} and \textbf{\textit{Iconicity}} features, as they are the most significant in identifying the form of gesture and their associated meaning.

\subsection{GloVe}

GloVe \cite{GloVe} is a fairly new model for word representation for
Global Vectors that  directly captures the global corpus statistics. Many Natural Language Processing (NLP) techniques such as GloVe, transform words into fixed dimensional vectors. These fixed dimensional vectors are learned from the co-occurrence relationship of words in large amounts of text. The learned vector representations of words help perform various NLP tasks, including word pair similarity, and named entity recognition. We use the GloVe model pre-trained on 6 Billion words from the 2014 Wikipedia dataset. 

In this work, we opted for GloVe model as it directly gives us a similarity score between two words. Newer word-vector representations like GPT-3 are not open source and only available on demand.  For this work, we use this pre-trained vector representation to calculate the pairwise similarity between the words in our dataset and the new technical terms. The cosine similarity metric is used for this purpose.







\section{Methodology}

\begin{figure*}
	\centering
	\includegraphics[width=\textwidth]{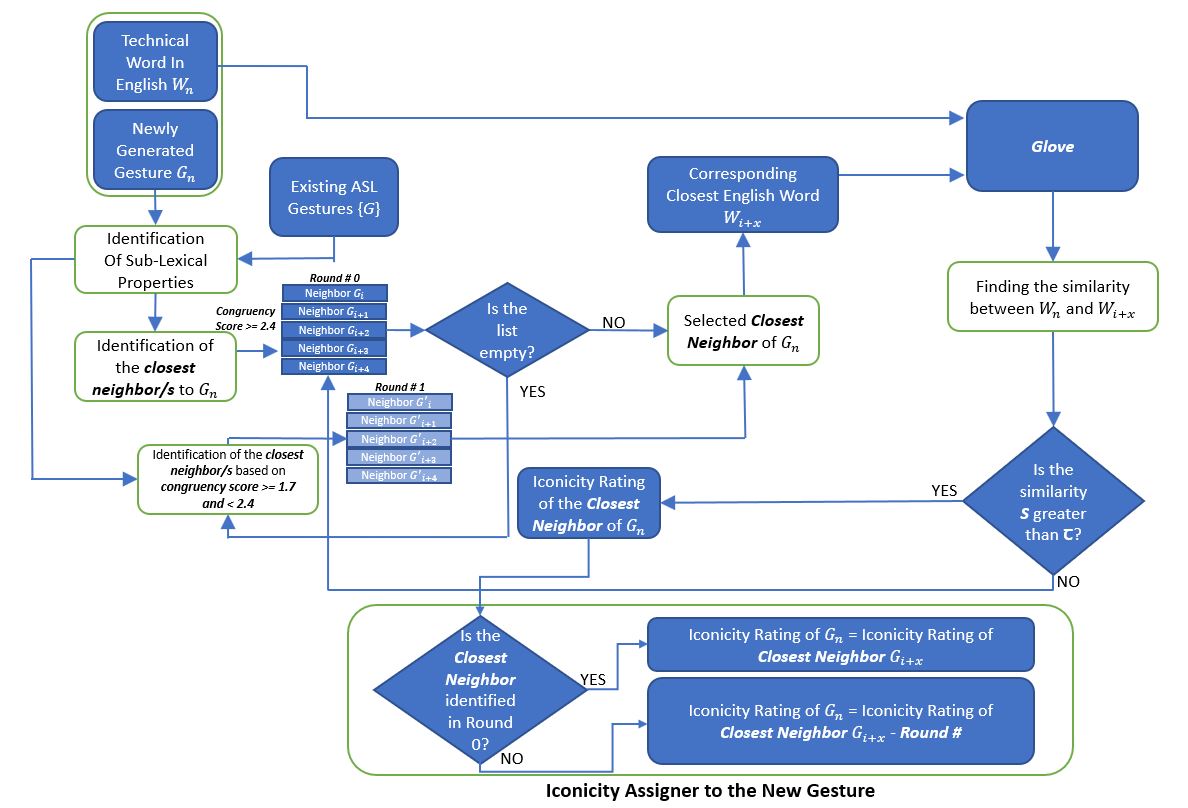}
	\caption{\textbf{\textit{EdGCon}}: Iconicty Rating Assigner for Newly Generated Gestures}
	\label{fig:EdGCon}
\end{figure*}

In this section, we discuss different components of \textbf{\textit{EdGCon}} (Figure \ref{fig:EdGCon}) and the proposed process.

\subsection{ASL Gesture Expression in terms of Sub-Lexical Properties}

\begin{figure}
	\centering
	\includegraphics[width=\columnwidth]{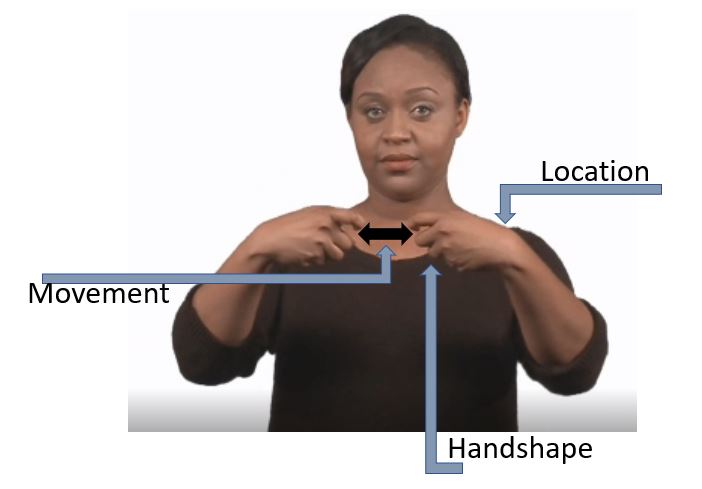}
	\caption{Sub-Lexical Properties of ASL~\cite{DeafTec}}
	\label{fig:ASL}
\end{figure}

One of the main components of our work is the expression of a gesture as a temporal sequence of \textbf{\textit{Sub-Lexical Properties}}. A \textbf{sub-lexical property} in a gesture can be defined as an indivisible component that has a meaningful association with an object, a body part, action, or a physical space in any signed language. In ASL, (Fig \ref{fig:ASL}) there are three sub-lexical properties related to \textit{handshape}, \textit{location} of the palm, and \textit{movement} of the palm in each hand. Any gesture in ASL can be expressed as a unique combination of these three lexicons. In ASL, commonality of these properties may also indicate similarity in meaning of the corresponding word.  

\begin{figure}
	\centering
	\includegraphics[width=\columnwidth]{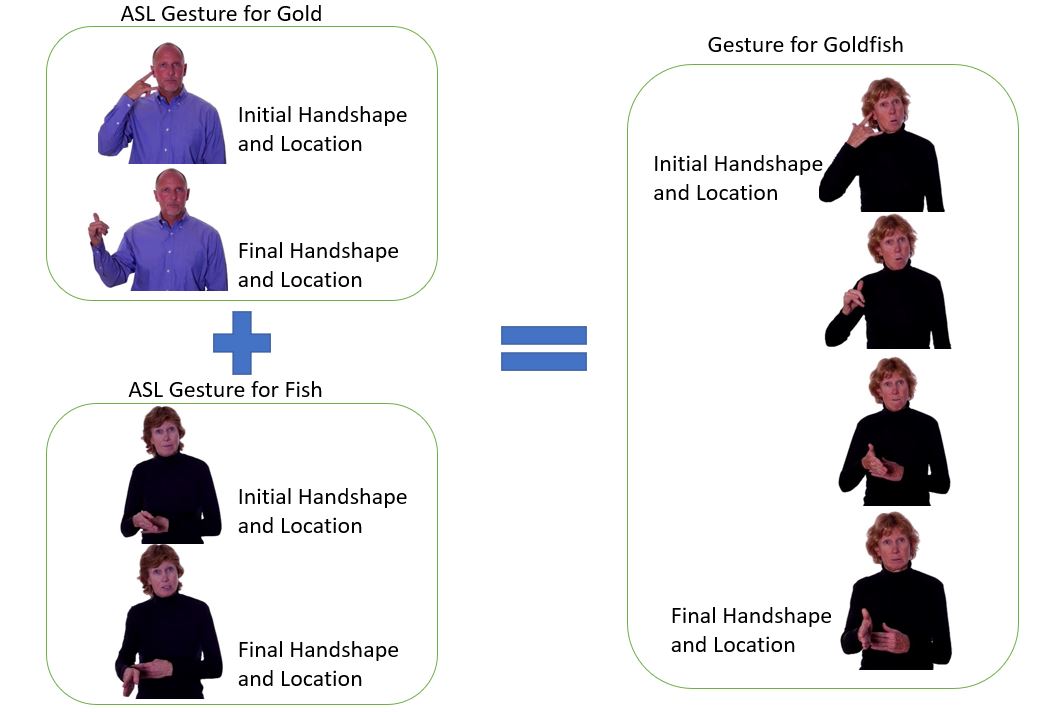}
	\caption{Gesture for Gold, Fish and Goldfish (collected from \textit{Signingsavvy.com} \cite{SignSavvy}) }
	\label{fig:Goldfish}
	
\end{figure}

For example, the ASL handshape used for ``Goldfish'' starts off with the handshape for ``Gold'' and then morphs into the handshape for ``Fish'' (as seen in Fig \ref{fig:Goldfish}). The handshape and movement for the gesture ``Father'' is same as that of ``Mother'', but the location for ``Father'' is near the forehead while that of ``Mother'' is near the chin. In fact, this difference in location often indicates the gender of the person in ASL words. Hence, the ability to express a gesture as a temporal sequence of the sub-lexical properties (handshape, location, and movement) helps in identifying the commonality in different gestures. We use these sub-lexical properties to identify similarities between these gesture forms and find how these similarities relate to the meaning of the words.

We build our \textit{Gesture Expression Set} based on the three sub-lexical properties of ASL gestures: 1) location, 2) movement and 3) handshape. Each gesture in ASL starts with an initial handshape, initial location and ends with a final handshape and final location. Between the initial handshape, location and final handshape and location, there is a specific movement. These three components are unique properties of a gesture since each of the handshapes, locations and movements provide gestural cues that allow ASL speakers to identify individual words.

We consider the \textit{Gesture Expression Set}, $\Gamma$, where $\Gamma$ = $\Gamma_{\it{H}} \, \bigcup \, \Gamma_{\it{L}} \, \bigcup \, \Gamma_{\it{M}}$. Here, $\Gamma_{\it{H}}$ is the set of handshapes, $\Gamma_{\it{L}}$ is the set of locations and $\Gamma_{\it{m}}$ is the set of movements. We define our \textit{Gesture Expressions} in terms of these sub-lexical properties. Representing expressions of gestures as a set of production rules in a context free grammar, $GE$:

\begin{scriptsize}
	\begin{eqnarray}
		\label{eqn:Grammar}
		&Handshapes (H)& \rightarrow \Gamma_{\it{H}}\\\nonumber
		&Locations (L)& \rightarrow \Gamma_{\it{L}}\\\nonumber
		&Movements (M)& \rightarrow  \Gamma_{\it{M}}\\\nonumber
		&GE& \rightarrow GE_{Left} GE_{Right}\\\nonumber
		&GE_{x}& \rightarrow H | \emptyset,\\\nonumber
		&where& x \in {Right, Left}   \\\nonumber
		&GE_{x}& \rightarrow H L\\\nonumber
		&GE_{x}& \rightarrow H L M H L
	\end{eqnarray}
\end{scriptsize}

Identification of sub-lexical properties is enhanced by utilizing this grammatical representation as discussed in \cite{Kamzin2020ICPR}.

\subsection{Identification of Sub-Lexical Properties}

We collect the results obtained from $Location$, $Handshape$ and $Movement$ recognition for automated identification of the sub-lexical properties. For recognition, $keypoints$ are obtained from each gesture execution. $Keypoints$ are the body parts that are tracked frame by frame throughout the video. Keypoint estimation is necessary to identify the location, movement and handshape of the gesture execution. $Keypoints$ for eyes, nose, shoulder, elbows, wrists and hands are collected using MediaPipe(\cite{MediaPipe}).

\textit{Location Recognition}: The automated recognition considers $start$ and $end$ locations of the hand position for pose estimation using MediaPipe \cite{MediaPipe}.  This model identifies wrist joint positions frame by frame from a video of ASL gesture execution in a 2D space for key points. The two axes namely $X$-axis (the line that connects the two shoulder joints) and $Y$-axis (perpendicular to the x-axis) are drawn based on the shoulders of the signer as a fixed reference. We divide the video canvas into 4 different sub-sections called buckets. Then, as the ASL user executes any given sign, the buckets are identified for the starting and ending location of the handshape. Location labels obtained from the system and gestures with common location for the start or the end of the handshape are considered to find the similarity. Later these labels are used to identify neighbors based on cosine similarity.

\textit{Handshape Recognition}: ASL signs differ in meaning based on the shape or orientation of the hands. We used MediaPipe to get the landmarks of both hands in every frame of the video. To ensure focused attention on the handshapes and eliminate the background noise, we draw the landmarks obtained for every frame on a corresponding black image. Once the landmarks are drawn on black background images for all frames, we extract the initial and final handshape for every gesture video. For this, the frames are divided in two halves. The second half’s median is considered as the final handshape frame. The first half is divided into two and the median frame of the first half’s second half is taken as an initial handshape frame. Once we have initial and final handshape frames for every gesture, we obtain the penultimate layer results using the pre-trained VGG16 model. The handshapes are then compared and matched based on their penultimate layer results’ cosine similarity.

\textit{Movement Recognition}: The hand movement recognition captures the movement of hands with respect to time from its start to the end. We used MediaPipe to get the pose estimation from every gesture. From that, both left and right wrists' key points are plotted as separate images for every gesture. After plotting the left and right wrists' key points for every gesture, we use the same process as that of handshape recognition by getting penultimate layer results and computing their cosine similarity.

\subsection{Auto-Assignment of Iconicity Rating}

Using \textbf{\textit{EdGCon}} (as seen in Fig. \ref{fig:EdGCon}), we propose a two step process to automatically assign iconicity ratings to newly generated gestures. The two steps are based on the two basic properties - \textit{sub-lexical properties} and \textit{iconicity}. 

\subsubsection{\textbf{Step 1: Identification of \textit{Closest Neighbors}}}
Each gesture essentially consists of two parts in its expression- \textit{The Gesture form, \textbf{G}}, and the \textit{Corresponding English word, \textbf{W}}. In this step, we identify the closest neighbor of the new gesture and find how closely related the respective words are in meaning. For any newly generated gesture, $G_n$ (with corresponding English word $W_n$), sub-lexical properties are identified and are matched with the sub-lexical properties of the existing ASL gestures in the DeafTec Data set $G$ (as seen in Fig. \ref{fig:EdGCon}). Based on the combined cosine similarity scores for location, movement and handshape, a \textit{\textbf{Congruency Score}} is calculated. Multiple rounds of neighbor identification are anticipated at this step and different levels of congruency score thresholds are determined for each round. A list of closest neighbors is collected in this round. If the collected list of neighbors in round\# 0 (as seen in Fig. \ref{fig:EdGCon}) is exhausted or empty, the neighbor list is collected from the next round. A lower congruency score threshold is determined for the next round and the congruency score is thus lowered for subsequent rounds.

From the collected list of neighbors, the topmost neighbor $G_i$ is selected with the corresponding English word $W_i$. Using $GloVe$, the similarity between $W_n$ and $W_i$ is computed. This similarity score, $S$, is passed on to the next step for \textit{Iconicity Rating Assignment}.

\subsubsection{\textbf{Step 2: \textit{Iconicity Rating} Assignment}}

For iconicity rating assignment, the similarity score from $GloVe$ is compared with a predetermined threshold. There are two possible outcomes of this comparison- 

a) The similarity score, $S$, is higher than the threshold $\tau$: this indicates that the identified neighbor gesture, $G_i$, is similar in the form of the new gesture, $G_n$, and the corresponding English words, $W_i$ and $W_n$ are also similar in meaning. In this case, we conclude that the new gesture is as iconic as the neighbor gesture and assigned the same iconicity rating as the neighbor to the new gesture.

b) The similarity score, $S$, is lower than the threshold $\tau$: This indicates that even though the identified neighbor gesture, $G_i$, is similar in the form of the new gesture, $G_n$, the corresponding English words, $W_i$ and $W_n$ are not similar in their meaning. In this case, we move on to the next neighbor in the collected list from Step 1 and repeat the process. 

If the neighbor list collected in round\# 0 gets exhausted or is empty, then the neighbor list is collected from round\# 1. In case that the closest neighbor gesture collected from round\# 1, $G'_i$, is similar in the form of the new gesture, $G_n$, and similarity score for the corresponding English words, $W_i$ and $W_n$ also passes the set threshold, the iconicity rating of the new gesture, $G_n$ will be ranked lower than the iconicity rating of $G'_i$. Since the similarities of the sub-lexical properties are based on a lower threshold for subsequent rounds (round\#1, round\#2, etc.). This lower ranking scoring is based on the round number that the neighbor list is collected from and is based on the following equation-
\begin{equation}
	\text{Iconicity Rating of } G_n = (\text{Iconicity Rating of } G_i - Round\#)
\end{equation}

In this work, we have used two rounds of neighbor identification based on a high and a low threshold, but subsequent rounds can be added with varying levels of congruency score.

\section{Evaluation \& Results}
In this section we discuss the data sets used, evaluation process and the results obtained using the proposed method. To evaluate \textit{\textbf{EdGCon}} and our proposed automated iconicity assignment process, we collected iconicity ratings assigned by \textit{\textbf{EdGCon}}, then compared them with the manual observation iconicity ratings to compute the accuracy.

\subsection{Data Sets}

We collected 31 technical gestures from DeafTEC STEM dictionary corpus \cite{DeafTec} and assigned iconicity ratings to them based on manual observation following the process described in ASL-LEX development \cite{ASL-LEX}. We used three hearing signers and the average of their iconicity ratings assigned to each gesture. We call this data set the DeafTec Data Set and considered it to be the ``Exisiting Gesture Set, G'' for the purpose of this work.

We then collected an additional 30 ad hoc and crowdsourced technical gestures from various resources on the internet. We assigned iconicity ratings based on manual observation to these gestures following the same previously mentioned process. This data set is called the ``New Technical Gesture Data Set''.
\begin{figure}
	\centering
	\includegraphics[width=\columnwidth]{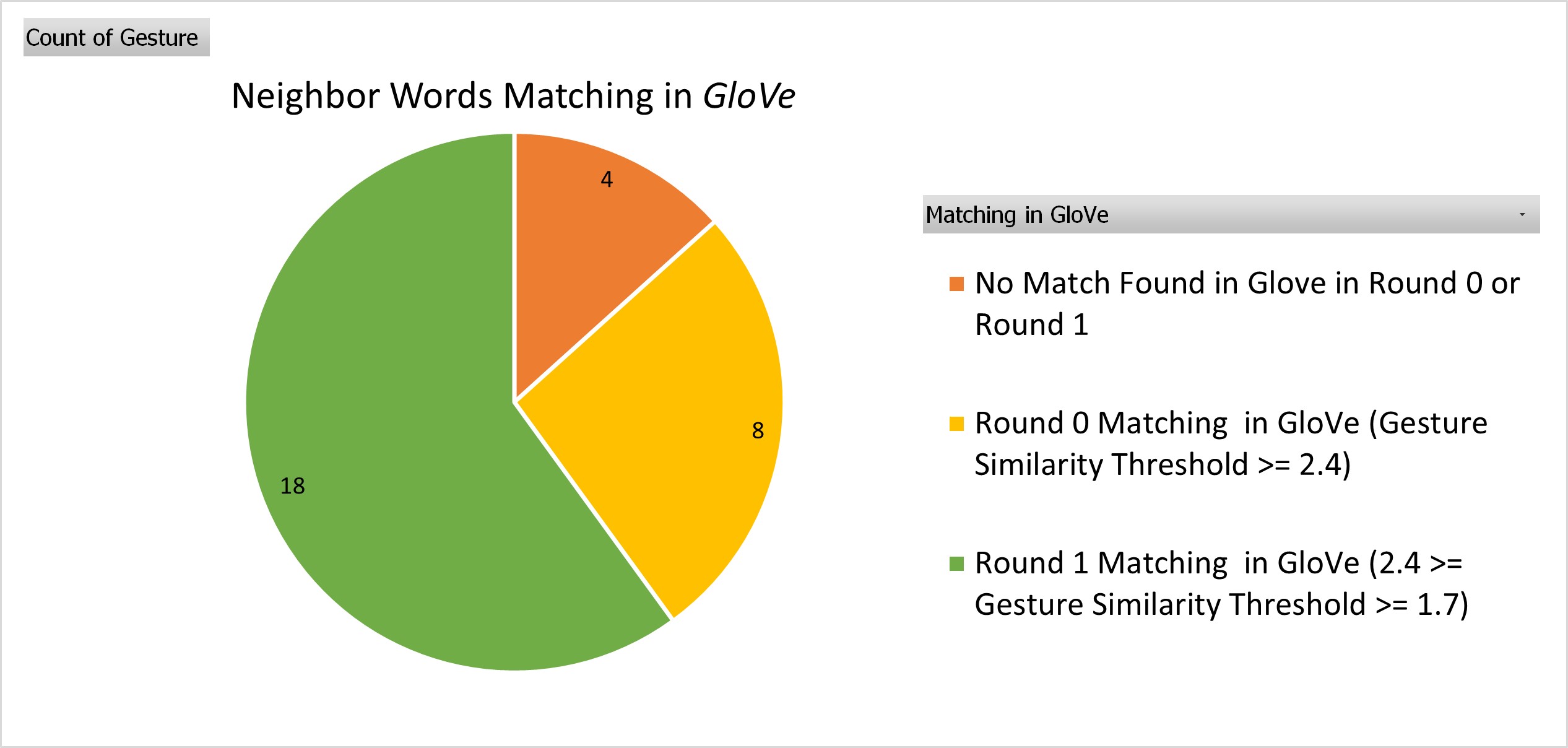}
	\caption{Neighbor Words in Round 0 and Round 1 that found matches in \textit{GloVe}}
	\label{fig:Rounds}
	
\end{figure}
\subsection{Evaluation}

We automatically identified the sub-lexical properties- handshape, location and movement, for the gestures in our existing data set and the gestures in our new technical gesture data set. For the 30 new technical gestures, we first identified the closest neighbors based on handshape identification selecting a 0.8 cosine similarity threshold. Since ASL gestures change meaning with different handshapes, we retrieved the location and movement cosine similarity measures for the neighbors identified in handshape recognition phase. We calculated the congruency scores for the identified neighbors by adding the cosine similarities for location, handshape and movement.

\begin{figure*}
	\centering
	\includegraphics[width=\textwidth]{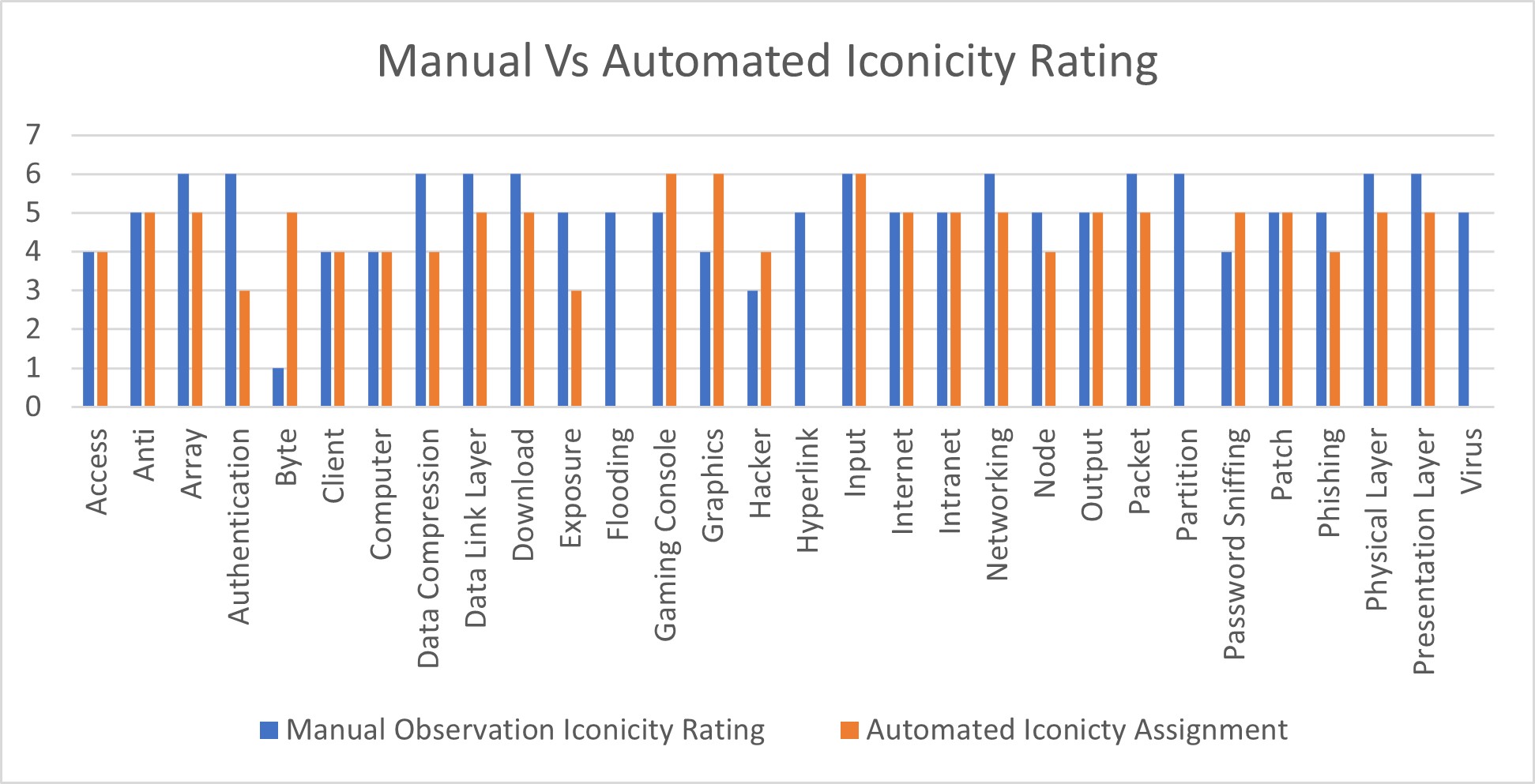}
	\caption{Manual Vs Automated Iconicity Ratings for the New Technical Gesture Set}
	\label{fig:ManualVAuto}
	
\end{figure*}

We have tested different scores and a congruency score threshold that produced best matching results was chosen. A congruency score threshold of 2.4 was set for \textit{\textbf{round\# 0}} and for each of the gestures ($G_n$) in the new technical gestures data set, a list of closest neighbors in gesture forms was created. For the first neighbor gesture in the list, $G_i$, with corresponding technical word $W_i$, we found the similarity score, $S$, between $W_n$ and $W_i$ using \textbf{\textit{GloVe}}. We set a threshold, $\tau$ of 0.3 to identify words that were similar in meaning. If $S$ was higher than $\tau$, then $EdGCon$ retrieved that iconicity rating for $G_i$ and assigned the same rating to $G_n$. If $S$ was lower than $\tau$, then the neighbor $G_{i+1}$ was chosen from the list created in \textit{\textbf{round\# 0}} and the process was repeated until an $S$ higher than $\tau$ was found or the list was exhausted (as seen in Fig. \ref{fig:EdGCon}). If the list was exhausted, the closest neighbor identification process was moved to \textit{\textbf{round\# 1}} with a lower congruency score to identify the neighbors  (2.4> congruency score >= 1.7 for \textit{\textbf{round\# 1}}, as seen in Fig. \ref{fig:EdGCon}) and previously described above mentioned process was repeated.

We collected the iconicity ratings assigned by \textit{\textbf{EdGCon}} to the new technical gesture set and computed the accuracy based on the manual iconicity ratings.

\subsection{Results}

\textit{\textbf{EdGCon}} was able to automatically assign iconicity ratings to a majority of the 30 new technical gestures based on our proposed process. For four gestures, \textit{\textbf{EdGCon}} was unable to find technical words with higher similarity scores than the set threshold even after second round of neighbor selection. For most of the remaining gestures, the automatic iconicity rating assignment was very close to the ratings assigned by manual observation (Fig. \ref{fig:ManualVAuto}). The automatic iconicity ratings of nine gestures were exactly same as the manual iconicity ratings, and the automatic ratings of twelve gestures were +1/-1 from the manual ratings.

For 8 gestures, a neighbor passing the \textit{\textbf{round\# 0}} threshold was found in \textit{\textbf{GloVe}} and for 18 gestures a neighbor passing the \textit{\textbf{round\# 1}} threshold was found in \textit{\textbf{GloVe}} (Figure \ref{fig:Rounds}). We excluded the four gestures for which no close neighbor word was identified based on the \textit{\textbf{round\# 0}} and \textit{\textbf{round\# 1}} thresholds in \textit{\textbf{GloVe}}. The total test words for accuracy calculation was set at 26. We also considered that the manual assignment of iconicity rating is subjective, was derived from a scale of 1 to 7, and based on a small number of observers. Thus, the automated rating of \textit{\textbf{EdGCon}} was set to be the same rating as manual if it was +1/-1 in distance in the 1 to 7 scale. \textit{\textbf{EdGCon}} was able to correctly assign \textit{\textbf{Iconicity Ratings}} to 21 of the 26 words. So, the accuracy of \textit{\textbf{EdGCon}} was computed to be 80.76\%.

\section{Use of EDGCon in DHH Education}

We present a usage scenario of the proposed tool, \textbf{\textit{EdGCon}}, to facilitate technical gesture generation. \textbf{\textit{EdGCon}} is envisioned as a part of \textit{CSignGen} framework (Fig \ref{fig:CSGen}) that has three components: a) a feedback driven gesture learning mechanism, which helps to learn a gesture, b) an iterative gesture generation mechanism, which helps to develop a unique gesture for a technical term that is lexically grounded, and c) a crowdsourcing platform to establish the acceptance of a gesture for a technical term as standard. \textit{Iconicity rating} is one of the properties that will determine whether a technical gesture can become a candidate for a standard gesture. 

\textit{CSignGen} will not only help DHH learners with a technical gesture learning platform, but can also facilitate the development of a corpus of lexically grounded technical gestures. This will not only help in recognizing complex concepts faster but will also help in sharing knowledge among DHH and hearing peers. \textit{CSignGen} utilizes a database of video examples of ASL concepts available in ASL online repositories such as SignSavvy~\cite{SignSavvy} and CS technical terms available on DeafTec website ~\cite{DeafTec}. When a Deaf individual or interpreter requires a technical word, they can utilize the search functionality of \textit{CSignGen} to search for a standard gesture. There are \textit{\textbf{three possible outcomes}} for this search: 

\begin{figure*}
	\centering
	\includegraphics[width=\textwidth]{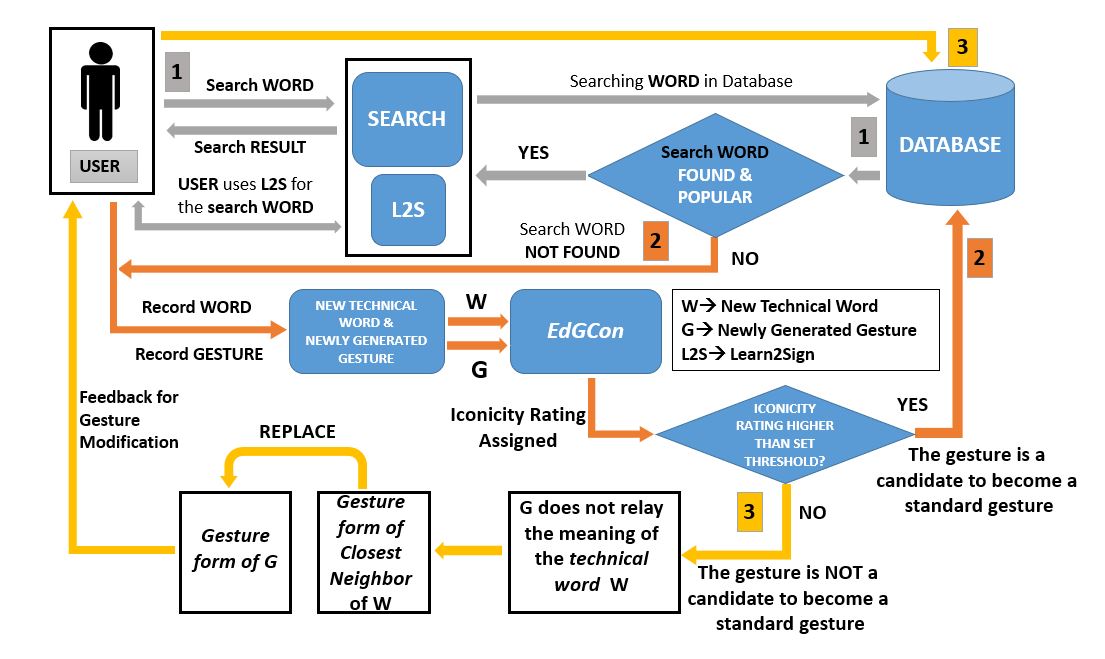}
	\caption{\textit{CSignGen} Framework for Gesture Learning, Generation \& Standardization}
	\label{fig:CSGen}
\end{figure*}

\noindent\textbf{\textit{a) No available gesture:}} The DHH individual and the interpreter will collaborate to develop a new gesture for the technical word, video record some demonstrations of the technical gesture, and will upload the videos to the \textit{CSignGen} server. \textbf{\textit{EdGCon}} will assign an \textit{Iconicty rating} to the new gesture and based on that rating the candidature of the technical gesture to be standardized will be determined.

\noindent\textbf{\textit{b) Few example gestures are available: }}The DHH student and the interpreter can either propose a new gesture (\textit{for this option, process in outcome (a) will be followed}), or choose from a small set of signs that are candidate signs. When a specific gesture reaches a threshold number of usages, it will be considered as the standard.

\noindent\textbf{\textit{c) Available standard gesture:}} The DHH student and the interpreter can use apps such as Learn2Sign (\textit{L2S})~\cite{learn2sign}, to learn the proper execution of the sign. 

Individually different components of the \textit{CSignGen} framework were evaluated in related research. The gesture learning mechanism was evaluated in \cite{BanerjeeAITutor}, and automated feedback based on sublexical properties was evaluated in \cite{HossainTwoStep2021}. A different approach to finding semantic similarity was investigated in \cite{HossainSemCog}. \textit{CSignGen} framework is a part of Computer Science Accessible Virtual Education (CSAVE) platform, a personalized technical education platform for DHH individuals \cite{Hossain2020PersonalizedTL}. Evaluation of the \textit{CSignGen} framework is not performed as part of this paper, but is included to illustrate the potential usage and significance of the tool proposed in the paper.

\section{Limitation \& Discussions}

With the small test data set of 30 gestures, and even with some of the gestures missing similar words in $GloVe$, the study shows that \textbf{\textit{EdGCon}} was able to correctly assign the iconicity ratings to 4/5 of the gestures. We expect the accuracy to be higher in a larger scale study with expanded data sets and domain specific word-vector model. In this section, we discuss some limitations of our evaluation of  \textbf{\textit{EdGCon}} and provide recommendations for future evaluations.

\noindent\textbf{\textit{Lack of Domain Specific Word-Vector Model}}: As mentioned previously, in \textit{\textbf{GloVe}} we were unable to find neighbors that pass the threshold for four words (Flooding, Hyperlink, Partition,\& Virus). \textit{\textbf{GloVe}} is a popular word vector representation model where words are placed in close proximity based on how frequently they appear together in the depository of text documents. While it is an excellent source for finding similarities between regular English words, most technical words fall in the same category in \textit{\textbf{GloVe}} as they frequently appear in the same technical document. A domain-specific word-vector model where technical words are categorized based on their function or purpose and then ranked on their frequency within the same document will help by significantly narrowing down the word list. This will help in improving identification of words that are both similar in gesture forms and closely related in their meanings.

\noindent\textbf{\textit{Few Available Gesture Repositories for Technical Words}}: We collected  data for our technical gesture corpus from the DeafTEC STEM dictionary, which is the pioneer in Deaf STEM education. However, many available technical words involved finger spelling of the word, which limited the application of this effort to assess similarity in gesture forms and their relation in meanings. After cleaning the dataset, excluding finger-spelled technical words, only 45 usable technical words were expressed in gestural form (\textit{of those, we have used 31 gestures for training our model and the remaining 14 along with 16 other crowdsourced gestures for testing the model}). There are other ad hoc technical gestures available in ASL crowd-sourcing websites, but there are no other STEM or technical specific gesture repositories. While this study presents one potential avenue to address this specific problem, large scale evaluations of the tool would require a larger technical gesture repository.

\section{Conclusion and Future Work}

This paper provides a tool to automatically assign iconicity ratings to newly generated gestures. Results obtained from the evaluation process shows that the automated iconicity assigner is able to accurately assign iconicity ratings for most of the gestures. The results obtained in this small-scale evaluation indicate that further investigation with a much larger data set is warranted with extensive comparison using different word-vector representation models. As shown through the $CSignGen$ usage scenario, \textbf{\textit{EdGCon}} can aid in generation and standardization of gestures in the technical domain and can also work as an important verification tool for crowd-sourced technical gestures. This can help enhance the development of expanded technical gesture databases and may facilitate DHH student participation in the technical field.

This work focused on the lexically grounded iconicity assignment process. An extensive evaluation of the tool with large data sets was not performed. In future, we intend to test the usage of \textbf{\textit{EdGCon}} tool as a part of $CSignGen$ framework with larger data sets of domain specific technical and non-technical words. The process presented in this paper can be implemented to assess the degree of iconicity for any new (or a less popular) gesture relative to existing gestures in the same domain. This process can be adopted in other gesture-based training such as dance performance, medical surgical precision, or military gesture learning. This process is a step towards automatic selection of appropriate words for new or uncommon gestures in automated continuous gesture translation. Automated continuous gesture translation can not only lead to improved participation in technical education, it can also facilitate future work spaces (\textit{like Amazon Fullfillment Centers}) where machines and humans will collaborate with each other, and machines can learn movements and processes from humans.

We have only focused on \textit{Iconicity} in this paper. Other gesture based comparison based on level of usage, idiomaticity etc. can be considered for future work in extending this research.


\bibliographystyle{ACM-Reference-Format} 
\bibliography{IAAI21,bib1,bib2014,bioheat,cite,cite1,CNSProposalMBE,FormalModel,impact,NSFSHB,IUI_version1,NSFSATC,RelatedWorks,sceptre,TECS}


\begin{thebibliography}{25}


\ifx \showCODEN    \undefined \def \showCODEN     #1{\unskip}     \fi
\ifx \showDOI      \undefined \def \showDOI       #1{#1}\fi
\ifx \showISBNx    \undefined \def \showISBNx     #1{\unskip}     \fi
\ifx \showISBNxiii \undefined \def \showISBNxiii  #1{\unskip}     \fi
\ifx \showISSN     \undefined \def \showISSN      #1{\unskip}     \fi
\ifx \showLCCN     \undefined \def \showLCCN      #1{\unskip}     \fi
\ifx \shownote     \undefined \def \shownote      #1{#1}          \fi
\ifx \showarticletitle \undefined \def \showarticletitle #1{#1}   \fi
\ifx \showURL      \undefined \def \showURL       {\relax}        \fi
\providecommand\bibfield[2]{#2}
\providecommand\bibinfo[2]{#2}
\providecommand\natexlab[1]{#1}
\providecommand\showeprint[2][]{arXiv:#2}

\bibitem[\protect\citeauthoryear{{ASL Clear}}{{ASL Clear}}{2020}]%
        {ASLClear}
\bibfield{author}{\bibinfo{person}{{ASL Clear}}.}
  \bibinfo{year}{2020}\natexlab{}.
\newblock \bibinfo{title}{The Learning Center for the Deaf}.
\newblock
\newblock
\urldef\tempurl%
\url{https://www.tlcdeaf.org/asl-clear}
\showURL{%
\tempurl}
\newblock
\shownote{Accessed: 2020-09-14}.


\bibitem[\protect\citeauthoryear{Banerjee, Lamrani, Hossain, Paudyal, and
  Gupta}{Banerjee et~al\mbox{.}}{2020}]%
        {BanerjeeAITutor}
\bibfield{author}{\bibinfo{person}{Ayan Banerjee}, \bibinfo{person}{Imane
  Lamrani}, \bibinfo{person}{Sameena Hossain}, \bibinfo{person}{Prajwal
  Paudyal}, {and} \bibinfo{person}{Sandeep K.~S. Gupta}.}
  \bibinfo{year}{2020}\natexlab{}.
\newblock \showarticletitle{AI Enabled Tutor for Accessible Training}. In
  \bibinfo{booktitle}{\emph{Artificial Intelligence in Education}}.
  \bibinfo{publisher}{Springer International Publishing},
  \bibinfo{address}{Cham}, \bibinfo{pages}{29--42}.
\newblock
\showISBNx{978-3-030-52237-7}


\bibitem[\protect\citeauthoryear{Boden and Rohlfing}{Boden and
  Rohlfing}{2021}]%
        {IconicityWordProduction}
\bibfield{author}{\bibinfo{person}{Ulrich~J. Boden} {and}
  \bibinfo{person}{Katharina Rohlfing}.} \bibinfo{year}{2021}\natexlab{}.
\newblock \showarticletitle{Progressive Reduction of Iconic Gestures
  Contributes to School-Aged Children’s Increased Word Production}.
\newblock \bibinfo{journal}{\emph{Frontiers in Psychology}}
  \bibinfo{volume}{12} (\bibinfo{date}{04} \bibinfo{year}{2021}),
  \bibinfo{pages}{651725}.
\newblock
\urldef\tempurl%
\url{https://doi.org/10.3389/fpsyg.2021.651725}
\showDOI{\tempurl}


\bibitem[\protect\citeauthoryear{Caselli, Sehyr, Cohen-Goldberg, and
  Emmorey}{Caselli et~al\mbox{.}}{2016}]%
        {ASL-LEX}
\bibfield{author}{\bibinfo{person}{Naomi Caselli}, \bibinfo{person}{Zed Sehyr},
  \bibinfo{person}{Ariel Cohen-Goldberg}, {and} \bibinfo{person}{Karen
  Emmorey}.} \bibinfo{year}{2016}\natexlab{}.
\newblock \showarticletitle{ASL-LEX: A lexical database of American Sign
  Language}.
\newblock \bibinfo{journal}{\emph{Behavior Research Methods}}
  \bibinfo{volume}{49} (\bibinfo{date}{05} \bibinfo{year}{2016}).
\newblock
\urldef\tempurl%
\url{https://doi.org/10.3758/s13428-016-0742-0}
\showDOI{\tempurl}


\bibitem[\protect\citeauthoryear{DeafTEC}{DeafTEC}{2020}]%
        {DeafTec}
\bibfield{author}{\bibinfo{person}{DeafTEC}.} \bibinfo{year}{2020}\natexlab{}.
\newblock \bibinfo{title}{DeafTEC STEM Sign Video Dictionary}.
\newblock
\newblock
\urldef\tempurl%
\url{https://deaftec.org/stem-dictionary/}
\showURL{%
\tempurl}
\newblock
\shownote{Accessed: 2020-09-14}.


\bibitem[\protect\citeauthoryear{Fellbaum}{Fellbaum}{2005}]%
        {WordNet}
\bibfield{author}{\bibinfo{person}{Christiane Fellbaum}.}
  \bibinfo{year}{2005}\natexlab{}.
\newblock \showarticletitle{WordNet and Wordnets}.
\newblock In \bibinfo{booktitle}{\emph{Encyclopedia of Language and
  Linguistics}}, \bibfield{editor}{\bibinfo{person}{Alex Barber}} (Ed.).
  \bibinfo{publisher}{Elsevier}, \bibinfo{pages}{2--665}.
\newblock


\bibitem[\protect\citeauthoryear{{Gallaudet University}}{{Gallaudet
  University}}{2012}]%
        {GD}
\bibfield{author}{\bibinfo{person}{{Gallaudet University}}.}
  \bibinfo{year}{2012}\natexlab{}.
\newblock \bibinfo{title}{Workshop for Emerging Deaf and Hard of Hearing
  Scientists}.
\newblock
\newblock


\bibitem[\protect\citeauthoryear{Hossain, Banerjee, and Gupta}{Hossain
  et~al\mbox{.}}{2020}]%
        {Hossain2020PersonalizedTL}
\bibfield{author}{\bibinfo{person}{Sameena Hossain}, \bibinfo{person}{Ayan
  Banerjee}, {and} \bibinfo{person}{Sandeep K.~S. Gupta}.}
  \bibinfo{year}{2020}\natexlab{}.
\newblock \showarticletitle{Personalized Technical Learning Assistance for Deaf
  and Hard of Hearing Students}. In \bibinfo{booktitle}{\emph{Workshop on
  Artificial Intelligence for Education, AAAI 2020}}. New York, New York, USA.
\newblock


\bibitem[\protect\citeauthoryear{Hossain, Banerjee, and Gupta}{Hossain
  et~al\mbox{.}}{2022}]%
        {HossainSemCog}
\bibfield{author}{\bibinfo{person}{Sameena Hossain}, \bibinfo{person}{Ayan
  Banerjee}, {and} \bibinfo{person}{Sandeep K.~S. Gupta}.}
  \bibinfo{year}{2022}\natexlab{}.
\newblock \showarticletitle{Quantifying Semantic Congruence to Aid in Technical
  Gesture Generation in Computing Education}.
  \bibinfo{publisher}{Springer-Verlag}, \bibinfo{address}{Berlin, Heidelberg},
  \bibinfo{pages}{329–333}.
\newblock
\showISBNx{978-3-031-11646-9}
\urldef\tempurl%
\url{https://doi.org/10.1007/978-3-031-11647-6_63}
\showDOI{\tempurl}


\bibitem[\protect\citeauthoryear{Hossain, Kamzin, Amperayani, Paudyal,
  Banerjee, and Gupta}{Hossain et~al\mbox{.}}{2021}]%
        {HossainTwoStep2021}
\bibfield{author}{\bibinfo{person}{Sameena Hossain}, \bibinfo{person}{Azamat
  Kamzin}, \bibinfo{person}{Venkata Naga Sai~Apurupa Amperayani},
  \bibinfo{person}{Prajwal Paudyal}, \bibinfo{person}{Ayan Banerjee}, {and}
  \bibinfo{person}{Sandeep K.~S. Gupta}.} \bibinfo{year}{2021}\natexlab{}.
\newblock \showarticletitle{Engendering Trust in Automated Feedback: A Two Step
  Comparison of Feedbacks in Gesture Based Learning}. In
  \bibinfo{booktitle}{\emph{Artificial Intelligence in Education}}.
  \bibinfo{publisher}{Springer International Publishing},
  \bibinfo{address}{Cham}, \bibinfo{pages}{190--202}.
\newblock
\showISBNx{978-3-030-78292-4}


\bibitem[\protect\citeauthoryear{{IES NCES}}{{IES NCES}}{2019}]%
        {NCES}
\bibfield{author}{\bibinfo{person}{{IES NCES}}.}
  \bibinfo{year}{2019}\natexlab{}.
\newblock \bibinfo{title}{Digest of Educational Statistics}.
\newblock
\newblock
\urldef\tempurl%
\url{https://nces.ed.gov/}
\showURL{%
\tempurl}


\bibitem[\protect\citeauthoryear{Kamzin, Amperayani, Sukhapalli, Banerjee, and
  Gupta}{Kamzin et~al\mbox{.}}{2021}]%
        {Kamzin2020ICPR}
\bibfield{author}{\bibinfo{person}{A Kamzin}, \bibinfo{person}{V.~N. S.~A.
  Amperayani}, \bibinfo{person}{P. Sukhapalli}, \bibinfo{person}{A. Banerjee},
  {and} \bibinfo{person}{S.~K.~S. Gupta}.} \bibinfo{year}{2021}\natexlab{}.
\newblock \showarticletitle{{Concept Embedding through Canonical Forms: A Case
  Study on Zero-Shot ASL Recognition}}. In \bibinfo{booktitle}{\emph{25th
  International Conference on Pattern Recognition. Forthcoming}}. Milan, Italy.
\newblock


\bibitem[\protect\citeauthoryear{Kandana~Arachchige, Simoes~Loureiro, Wivine,
  Rossignol, and Lefebvre}{Kandana~Arachchige et~al\mbox{.}}{2021}]%
        {IconicitySpeechComprehension}
\bibfield{author}{\bibinfo{person}{Kendra Kandana~Arachchige},
  \bibinfo{person}{Isabelle Simoes~Loureiro}, \bibinfo{person}{Blekic Wivine},
  \bibinfo{person}{Mandy Rossignol}, {and} \bibinfo{person}{Laurent Lefebvre}.}
  \bibinfo{year}{2021}\natexlab{}.
\newblock \showarticletitle{The Role of Iconic Gestures in Speech
  Comprehension: An Overview of Various Methodologies}.
\newblock \bibinfo{journal}{\emph{Frontiers in Psychology}}
  \bibinfo{volume}{12} (\bibinfo{date}{04} \bibinfo{year}{2021}),
  \bibinfo{pages}{1--15}.
\newblock
\urldef\tempurl%
\url{https://doi.org/10.3389/fpsyg.2021.634074}
\showDOI{\tempurl}


\bibitem[\protect\citeauthoryear{Liddell}{Liddell}{2003}]%
        {LiddelASLGrammar}
\bibfield{author}{\bibinfo{person}{Scott Liddell}.}
  \bibinfo{year}{2003}\natexlab{}.
\newblock \showarticletitle{Grammar, Gesture, and Meaning in American Sign
  Language}.
\newblock \bibinfo{journal}{\emph{Grammar, Gesture, and Meaning in American
  Sign Language}} (\bibinfo{date}{03} \bibinfo{year}{2003}).
\newblock
\showISBNx{9780521816205}


\bibitem[\protect\citeauthoryear{Lualdi, Hudson, Fellbaum, and Buchholz}{Lualdi
  et~al\mbox{.}}{2019}]%
        {ASLNet2019}
\bibfield{author}{\bibinfo{person}{Colin Lualdi}, \bibinfo{person}{Jack
  Hudson}, \bibinfo{person}{Christiane Fellbaum}, {and} \bibinfo{person}{Noah
  Buchholz}.} \bibinfo{year}{2019}\natexlab{}.
\newblock \showarticletitle{Building {ASLN}et, a {W}ordnet for {A}merican
  {S}ign {L}anguage}. In \bibinfo{booktitle}{\emph{Proceedings of the 10th
  Global Wordnet Conference}}. \bibinfo{publisher}{Global Wordnet Association},
  \bibinfo{address}{Wroclaw, Poland}, \bibinfo{pages}{315--322}.
\newblock


\bibitem[\protect\citeauthoryear{Lugaresi, Tang, Nash, McClanahan, Uboweja,
  Hays, Zhang, Chang, Yong, Lee, Chang, Hua, Georg, and Grundmann}{Lugaresi
  et~al\mbox{.}}{2019}]%
        {MediaPipe}
\bibfield{author}{\bibinfo{person}{Camillo Lugaresi}, \bibinfo{person}{Jiuqiang
  Tang}, \bibinfo{person}{Hadon Nash}, \bibinfo{person}{Chris McClanahan},
  \bibinfo{person}{Esha Uboweja}, \bibinfo{person}{Michael Hays},
  \bibinfo{person}{Fan Zhang}, \bibinfo{person}{Chuo-Ling Chang},
  \bibinfo{person}{Ming Yong}, \bibinfo{person}{Juhyun Lee},
  \bibinfo{person}{Wan-Teh Chang}, \bibinfo{person}{Wei Hua},
  \bibinfo{person}{Manfred Georg}, {and} \bibinfo{person}{Matthias Grundmann}.}
  \bibinfo{year}{2019}\natexlab{}.
\newblock \showarticletitle{MediaPipe: A Framework for Perceiving and
  Processing Reality}. In \bibinfo{booktitle}{\emph{Third Workshop on Computer
  Vision for AR/VR at IEEE Computer Vision and Pattern Recognition (CVPR)
  2019}}.
\newblock
\urldef\tempurl%
\url{https://mixedreality.cs.cornell.edu/s/NewTitle_May1_MediaPipe_CVPR_CV4ARVR_Workshop_2019.pdf}
\showURL{%
\tempurl}


\bibitem[\protect\citeauthoryear{Ozcaliskan, Adamson, Dimitrova, and
  Baumann}{Ozcaliskan et~al\mbox{.}}{2017}]%
        {GestureDownSyndrome}
\bibfield{author}{\bibinfo{person}{Seyda Ozcaliskan}, \bibinfo{person}{Lauren
  Adamson}, \bibinfo{person}{Nevena Dimitrova}, {and}
  \bibinfo{person}{Stephanie Baumann}.} \bibinfo{year}{2017}\natexlab{}.
\newblock \showarticletitle{Early Gesture Provides a Helping Hand to Spoken
  Vocabulary Development for Children with Autism, Down Syndrome and Typical
  Development}.
\newblock \bibinfo{journal}{\emph{Journal of Cognition and Development}}
  \bibinfo{volume}{18} (\bibinfo{date}{06} \bibinfo{year}{2017}).
\newblock
\urldef\tempurl%
\url{https://doi.org/10.1080/15248372.2017.1329735}
\showDOI{\tempurl}


\bibitem[\protect\citeauthoryear{Paudyal, Lee, Kamzin, Soudki, Banerjee, and
  Gupta}{Paudyal et~al\mbox{.}}{2019}]%
        {learn2sign}
\bibfield{author}{\bibinfo{person}{Prajwal Paudyal}, \bibinfo{person}{Junghyo
  Lee}, \bibinfo{person}{Azamat Kamzin}, \bibinfo{person}{Mohamad Soudki},
  \bibinfo{person}{Ayan Banerjee}, {and} \bibinfo{person}{Sandeep~K.S. Gupta}.}
  \bibinfo{year}{2019}\natexlab{}.
\newblock \showarticletitle{Learn2Sign: Explainable AI for Sign Language
  Learning}.
\newblock \bibinfo{journal}{\emph{Explainable Smart Systems workshop in ACM
  Intelligent User Interfaces conference}} (\bibinfo{year}{2019}).
\newblock


\bibitem[\protect\citeauthoryear{Pennington, Socher, and Manning}{Pennington
  et~al\mbox{.}}{2014}]%
        {GloVe}
\bibfield{author}{\bibinfo{person}{Jeffrey Pennington},
  \bibinfo{person}{Richard Socher}, {and} \bibinfo{person}{Christopher~D.
  Manning}.} \bibinfo{year}{2014}\natexlab{}.
\newblock \showarticletitle{GloVe: Global Vectors for Word Representation}. In
  \bibinfo{booktitle}{\emph{Empirical Methods in Natural Language Processing
  (EMNLP)}}. \bibinfo{pages}{1532--1543}.
\newblock
\urldef\tempurl%
\url{http://www.aclweb.org/anthology/D14-1162}
\showURL{%
\tempurl}


\bibitem[\protect\citeauthoryear{Savvy}{Savvy}{2018}]%
        {SignSavvy}
\bibfield{author}{\bibinfo{person}{Signing Savvy}.}
  \bibinfo{year}{2018}\natexlab{}.
\newblock \bibinfo{title}{{Signing Savvy: Your Sign Language Resouce}}.
\newblock \bibinfo{howpublished}{\url{https://www.signingsavvy.com/}}.
\newblock
\newblock
\shownote{[Online; accessed 28-September-2018]}.


\bibitem[\protect\citeauthoryear{Stokoe}{Stokoe}{2003}]%
        {StokoeStructure}
\bibfield{author}{\bibinfo{person}{W.C.Jr Stokoe}.}
  \bibinfo{year}{2003}\natexlab{}.
\newblock \showarticletitle{Sign Language Structure}.
\newblock \bibinfo{journal}{\emph{Annual Review of Anthropology}}
  \bibinfo{volume}{9} (\bibinfo{date}{11} \bibinfo{year}{2003}),
  \bibinfo{pages}{365--390}.
\newblock


\bibitem[\protect\citeauthoryear{Stokoe, Casterline, and Croneberg}{Stokoe
  et~al\mbox{.}}{1976}]%
        {stokoe1976dictionary}
\bibfield{author}{\bibinfo{person}{W.C. Stokoe}, \bibinfo{person}{D.C.
  Casterline}, {and} \bibinfo{person}{C.G. Croneberg}.}
  \bibinfo{year}{1976}\natexlab{}.
\newblock \bibinfo{booktitle}{\emph{A Dictionary of American Sign Language on
  Linguistic Principles}}.
\newblock \bibinfo{publisher}{Linstok Press}.
\newblock
\showLCCN{76365702}
\urldef\tempurl%
\url{https://books.google.com/books?id=WjAFAQAAIAAJ}
\showURL{%
\tempurl}


\bibitem[\protect\citeauthoryear{Stokoe, for Linguistics, and for
  Applied~Linguistics}{Stokoe et~al\mbox{.}}{1970}]%
        {stokoe1970study}
\bibfield{author}{\bibinfo{person}{W.C. Stokoe},
  \bibinfo{person}{ERIC~Clearinghouse for Linguistics}, {and}
  \bibinfo{person}{Center for Applied~Linguistics}.}
  \bibinfo{year}{1970}\natexlab{}.
\newblock \bibinfo{booktitle}{\emph{The Study of Sign Language}}.
\newblock \bibinfo{publisher}{ERIC Clearinghouse for Linguistics, Center for
  Applied Linguistics}.
\newblock
\showLCCN{72025232}
\urldef\tempurl%
\url{https://books.google.com/books?id=l4RCAAAAIAAJ}
\showURL{%
\tempurl}


\bibitem[\protect\citeauthoryear{Stokoe}{Stokoe}{2005}]%
        {stokoe2005sign}
\bibfield{author}{\bibinfo{person}{William~C Stokoe}.}
  \bibinfo{year}{2005}\natexlab{}.
\newblock \showarticletitle{Sign language structure: An outline of the visual
  communication systems of the American deaf}.
\newblock \bibinfo{journal}{\emph{Journal of deaf studies and deaf education}}
  \bibinfo{volume}{10}, \bibinfo{number}{1} (\bibinfo{year}{2005}),
  \bibinfo{pages}{3--37}.
\newblock


\bibitem[\protect\citeauthoryear{Walter}{Walter}{2010}]%
        {NITDRIT}
\bibfield{author}{\bibinfo{person}{Gerard~G. Walter}.}
  \bibinfo{year}{2010}\natexlab{}.
\newblock \bibinfo{title}{Deaf and Hard of Hearing Students in Transition:
  Demographics with an Emphasis on STEM Education}.
\newblock
\newblock
\urldef\tempurl%
\url{https://www.ntid.rit.edu/sites/default/files/cat/Transition%20demographic%20report%206-1-10.pdf}
\showURL{%
\tempurl}


\end{thebibliography}

\end{document}